\documentclass[twocolumn,showpacs,preprintnumbers,amsmath,amssymb]{revtex4}
\usepackage{graphicx}
\begin{document}

\title{Dynamical description of vesicle growth and shape change}

\author{Richard G.~Morris$^{1}$, Duccio Fanelli$^{2}$ and 
Alan J.~McKane$^{1,2}$}

\affiliation{$^{1}$Theoretical Physics, School of Physics and Astronomy,
University of Manchester, Manchester M13 9PL, United Kingdom \\
$^{2}$Dipartimento di Energetica, University of Florence and INFN, 
Via S. Marta 3, 50139 Florence, Italy
}

\begin{abstract}
We systematize and extend the description of vesicle growth and shape change
using linear nonequilibrium thermodynamics. By restricting the study to 
shape changes from spheres to axisymmetric ellipsoids, we are able to give a 
consistent formulation which includes the lateral tension of the vesicle 
membrane. This allows us to generalize and correct a previous calculation.
Our present calculations suggest that, for small growing vesicles, a prolate 
ellipsoidal shape should be favored over oblate ellipsoids, whereas for large 
growing vesicles oblates should be favored over prolates. The validity of this 
prediction is examined in the light of the various assumptions made in its 
derivation.
\end{abstract}


\maketitle

\section{Introduction}
\label{intro}
Vesicles are fascinating structures for several reasons~\cite{lui06}: they 
occur in a wide range of shapes and sizes, they are cell-like and are 
frequently used in the modeling of proto-cells and they present a formidable 
task for the mathematical modeler. It is not hard to see why they are so 
difficult to describe theoretically. They are in effect a closed membrane 
made up of a lipid bilayer, which allows water and solutes to permeate from 
the environment into the interior of the vesicle~\cite{lui06,sei97}. While 
they are frequently modeled as a two-dimensional surface, the finite 
thickness of the bilayer as well as its structure, plays an important role in 
understanding their behavior~\cite{sei97}. In a pioneering work, 
Helfrich~\cite{hel73,oyy99} used the analogy between the rod-like shape of 
nematic liquid crystals and the lipids forming the vesicle bilayer, to write 
down an expression for the energy of the membrane. He envisaged the membrane 
as a curved surface, and accounted for the fact that the membrane had 
structure through the introduction of a phenomenological constant, $C_0$, 
the spontaneous curvature. Since then a variety of other models have been 
proposed, most of which start from the idea of a purely geometrical surface 
to represent the membrane, with additional structure introduced in various 
ways~\cite{sei97}.

For most of the period since these models were proposed the main focus of
their study has been to look for the shapes with the smallest energy under 
fixed conditions of constant volume, $V$ and constant surface area, $A$. It 
was argued that this would give the expected shape of the vesicle at these 
values of $V$ and $A$, and indeed a range of shapes emerged from this 
analysis~\cite{sei97}. However this is a purely static approach --- there is 
no mechanism which postulates how the transitions between different shapes 
occurred, or how long these changes take. We recently carried out a 
preliminary study of a dynamical process designed to describe the change in 
vesicle shape~\cite{fan08}. The change in surface area occurred because of 
the slow accretion of lipids onto the surface from the fluid surrounding the 
vesicle. This resulted in a change in the volume of the vesicle due to the
influx of water and solutes through the membrane. The process was assumed to be 
sufficiently slow that the formalism of linear nonequilibrium thermodynamics 
(LNET) could be used~\cite{deG84}. The expectation was to find stability 
conditions which determine the shape of the vesicle at various stages of its 
growth.

While we believe that modeling vesicles as surfaces with an energy of the 
Helfrich type, which change their shape and size according to LNET, is a 
minimalist description which is capable of answering many interesting 
questions, it is still extremely mathematically complex. The surface may be 
of an arbitrary shape, and so requires the formalism of differential geometry 
for its study, the actual dynamics, as opposed to the statics, of mixtures 
of the rod-like lipids and the point-like fluid molecules is very non-trivial, 
and the nonequilibrium thermodynamics of discontinuous structures such as 
vesicles as been little studied because of its difficulty. Some of these 
questions will be discussed in a forthcoming publication~\cite{mor10}, where 
we will develop the formalism in a systematic fashion. However in our view it 
is also valuable to proceed in stages, and build on what has gone
before~\cite{fan08,boz04,boz07,boz09,fan09}, and not to attempt to introduce 
all aspects of the problem simultaneously. This is the philosophy that we will 
follow in this paper; we will extend our previous calculation to include 
surface tension (frequently called lateral tension when discussing vesicles) 
in the context of axisymmetric ellipsoidal shapes, treating the area $A$, as 
well as the volume, $V$, as thermodynamic variables. This will form the basis 
for the treatment of more general shapes in the future. We will also correct 
some of the analysis presented in~\cite{fan08}. 

\section{Formalism}
\label{formalism}
We shall in common with most other authors adopt the spontaneous curvature 
model of Helfrich~\cite{hel73} in which the energy of the membrane is given
by 
\begin{equation}
E_{m} = \frac{\kappa}{2} \oint_{A} \left( 2H - C_{0} \right)^{2}\,\mathrm{d}A\,,
\label{mem_ener}
\end{equation}
where $H$ is the local mean curvature, $\mathrm{d}A$ is an element of the 
surface $A$ and $\kappa$ is the bending rigidity. Since a purely static 
analysis of Eq.~(\ref{mem_ener}) shows that when a spherical surface becomes 
unstable it is replaced with an ellipsoid~\cite{ou87-89}, we will restrict 
ourselves to shapes which are spheres or axisymmetric ellipsoids. These will be 
parametrized in Cartesian coordinates by
\begin{eqnarray}
x &=& a \sin \theta \cos \phi\,, \nonumber \\
y &=& a \sin \theta \sin \phi\,, \nonumber \\
z &=& c \cos \theta\,,
\label{coords}
\end{eqnarray}
where $0 \leq \phi < 2\pi$, $0 \leq \theta \leq \pi$ and where $a$ and $c$ 
are constants. For a sphere $a=c\equiv r$, the radius. We have not included 
a term proportional to the Gaussian curvature in the expression for the energy,
since this is a constant for the kinds of surfaces that we are considering 
here~\cite{sei97,oyy99}.

The surface area, volume and membrane energy (\ref{mem_ener}) can all be 
found in closed form for an axisymmetric ellipsoid and are given explicitly
in the appendix of \cite{fan08} (there is however a typographical error
in Eqs.~(A9) of that paper: the minus sign in the factor $-2a^{3}/3c^{4}$ should
be absent). They are all functions of the two variables $a$ and $c$ which 
characterize the ellipsoid, with $E_{m}$ additionally depending on the 
parameters $\kappa$ and $C_0$. Since the volume takes the simple form 
$V = 4\pi a^{2}c/3$, $c$ may be explicitly eliminated in favor of $a$ and $V$ 
in the expressions for $A$ and $E_{m}$. Subsequently $A=A(V,a)$ may be inverted
to give $a=a(V,A)$. In this way we see that the energy of the membrane is a 
function of $V$ and $A$ only. Since in the Helfrich model the membrane is 
completely described by this energy, we deduce that the thermodynamic 
description of the membrane can be achieved using these two variables.

As previously indicated we use LNET to describe the time evolution of 
the vesicle. Throughout, we assume that the temperature, $T$, is fixed, 
since we are not interested in changes in shape which come about because 
of a change in temperature. For simplicity, we assume that no solutes are 
present, although it is not too difficult to extend the treatment to include 
them.

The system is assumed to comprise two regions, the \textit{interior} and 
\textit{exterior}, that are separated by a third region, the \textit{membrane}.
This third region is assumed to be very thin, and our aim is to modify 
the thermodynamic description so that the membrane can (following Helfrich) 
be simply regarded as a boundary, or geometrical surface, between the first
two regions. In the formalism of LNET, each region is assumed to be in local 
equilibrium~\cite{deG84}, with the thermodynamic relation 
$T\mathrm{d}S=\mathrm{d}E - \mathrm{d}W$ holding in each region. Here $S$ is 
the entropy, $E$ the internal energy and $W$ the work done on the 
system~\cite{adk83}. The three regions will be labeled $i$, $e$ and $m$ 
respectively.

The internal energy and work done have the following forms:
\begin{itemize}
\item[(i)] The internal energy of the interior, $E_i$, and exterior, $E_e$, will
be that of the fluid in these two regions, in our case water. Their sum is 
denoted by $E_{w}$. The internal energy of the membrane is given by 
Eq.~(\ref{mem_ener}).
\item[(ii)] Considering the interior and exterior regions individually, each
will have work done on them of $-P\mathrm{d}V$ if their volumes change, where 
$P$ is the pressure in that region. Adding these gives the total contribution 
for work of this type to be $-P_{i}\mathrm{d}V_{i}-P_{e}\mathrm{d}V_{e}$. The 
membrane will be assumed to have negligible volume, and so gives no 
contribution. This also means that 
$\mathrm{d}V_{i}=-\mathrm{d}V_{e}\equiv \mathrm{d}V$, and so the total work 
done on the system due to the interior increasing its volume by $\mathrm{d}V$ 
is $(P_{e}-P_{i})\mathrm{d}V$. 
\item[(iii)] There will also be a work done if the area of the membrane 
increases by $\mathrm{d}A$, equal to $\sigma \mathrm{d}A$, where $\sigma$ is 
the surface tension~\cite{adk83}. This would exist even if the system 
consisted of two regions of different fluids, with no membrane separating them.
The existence of a membrane with non-trivial structure separating the interior 
and exterior, means that $\sigma$ will have a more complex functional form 
which reflects this structure. For this reason we will follow the usual usage 
in this field and refer to it as the lateral tension.
\end{itemize}

Adding up the contributions from the three regions one finds
\begin{equation}
T\mathrm{d}S = \mathrm{d}E_{w} + \mathrm{d}E_{m} - (\Delta P)\,\mathrm{d}V - 
\sigma\,\mathrm{d}A,
\label{thermo}
\end{equation}
where now $S$ is the total entropy of the system and 
$\Delta P \equiv (P_{e}-P_{i})$ is the pressure difference between the exterior
and interior. The membrane does not explicitly appear in the terms relating to
work. We can also eliminate it from the internal energy by noting that, since 
$E_{m}$ is a function of $V$ and $A$,
\begin{equation}
\mathrm{d}E_{\rm m} = \left( \frac{\partial E_{m}}{\partial V} \right)_{A} 
\mathrm{d}V + \left( \frac{\partial E_{m}}{\partial A} \right)_{V} \mathrm{d}A.
\label{differential}
\end{equation}
This allows us to write the thermodynamic relation (\ref{thermo}) for the 
system as
\begin{equation}
T\mathrm{d}S = \mathrm{d}E_{w} - \left( \Delta P \right)_{\rm eff}\,\mathrm{d}V 
- \left( \sigma_{\rm eff}\right)\,\mathrm{d}A,
\label{therm_rel}
\end{equation}
where
\begin{equation}
\left( \Delta P \right)_{\rm eff} = \Delta P -
\left( \frac{\partial E_{m}}{\partial V} \right)_{A} \ ; \
\sigma_{\rm eff} = \sigma - \left( \frac{\partial E_{m}}{\partial A} \right)_{V}.
\label{def_eff}
\end{equation}
Equation (\ref{therm_rel}) is the thermodynamic relation for two regions 
separated by a boundary with no material properties. The effect of the 
membrane simply changes the pressure difference and lateral tension from 
$\Delta P$ to $(\Delta P)_{\rm eff}$ and from $\sigma$ to $\sigma_{\rm eff}$ 
respectively. Therefore, as long as we make these replacements, we may ignore 
the membrane from a thermodynamic point of view, and simply treat it as a 
boundary which separates the inside of the vesicle from the environment. In 
our previous analysis~\cite{fan08} we did not treat the area as an independent 
variable. This is clearly consistent for spherical vesicles, but not for those 
which have an axisymmetric ellipsoidal shape. 

\section{Dynamics}
\label{dyn}
One of the central features of LNET is the relation between the forces, 
$X_{a}$, which cause the state of the system to change, and the fluxes, $J_a$, 
which are the result of these changes~\cite{deG84}. Within the formalism of
LNET these are linearly related: $J_{a} = \sum_{b} L_{ab} X_{b}$, where the
$L_{ab}$ are constants, the Onsager coefficients. The forces and fluxes can be 
identified in a systematic way~\cite{mor10}, but for the specific problem of 
interest to us here, where the relevant thermodynamic variables are simply 
$V$ and $A$, we may proceed more directly. We will also restrict ourselves 
to just one of the fluxes: that due to water flowing through the membrane 
into the interior of the vesicle. This will be denoted by $J_{w}$. The direct 
effect that causes this flux of water is the pressure difference between the 
exterior and interior regions~\cite{ked58,ked63}, which incorporating the 
effect of the membrane is $(\Delta P)_{\rm eff}$.

If we only took into account this direct effect, as we did in our previous 
treatment~\cite{fan08}, we would write $J_{w} = L_{p} (\Delta P)_{\rm eff}$, 
where $L_{p}$ is the hydraulic conductivity of the membrane. However there 
will also be an indirect effect~\cite{deG84} for which the driver will be 
$\sigma_{\rm eff}$, and so 
\begin{equation}
J_{w} = L_{p} (\Delta P)_{\rm eff} + L_{\sigma} \sigma_{\rm eff},
\label{J_w}
\end{equation}
where $L_{\sigma}$ is a second Onsager coefficient. We shall justify the choices
of forces and fluxes in more detail elsewhere~\cite{mor10}, but we can give
a simple microscopic argument showing how the flux proportional to the lateral
tension comes about. For a positively curved membrane, the lipids are arranged 
in a funnel-like configuration which inhibits the flow of water across the 
membrane. Decreasing the curvature of a membrane at any point therefore 
permits a greater flow of water due to the lipids becoming increasingly 
parallel. In this way, for fixed shapes, an increase in the surface area 
causes a slight alignment of adjacent lipids and hence permits a greater flow
across the membrane. This resultant flow is proportional to the lateral
tension, the measure of how the membrane energy changes with area.

We are now in a position to introduce the dynamics explicitly. The rate of 
increase of the volume of the vesicle will be given by 
$\mathrm{d}V/\mathrm{d}t = AJ_{w}$, and
so from Eqs.~(\ref{def_eff}) and (\ref{J_w}) we have
\begin{equation}
\frac{\mathrm{d}V}{\mathrm{d}t}=A\left\{L_{p}
\left[\Delta P-\left(\frac{\partial E_{m}}{\partial V}\right)_{A}\right] +
L_{\sigma}\left[\sigma-\left(\frac{\partial E_{m}}
{\partial A}\right)_{V}\right]\right\}
\label{dV_by_dt}
\end{equation}
We also need to describe how the surface area of the vesicle grows due to 
the inclusion of lipids from the environment. The simplest assumption is that 
these attach themselves uniformly over the entire surface, so that the
surface grows at a constant rate per unit area which we will denote by
$\lambda$~\cite{boz04,fan08}:
\begin{equation}
\frac{\mathrm{d}A}{\mathrm{d}t} = \lambda A\ \ \Rightarrow \ \ A(t) = 
A(0) e^{\lambda t}\,.
\label{growth_law}
\end{equation}

Before we go on to investigate the dynamics for an ellipsoidal shape, let us
briefly consider the result for a sphere. In this case $V$ and $A$ are not
independent variables, and so the thermodynamic relation (\ref{therm_rel}) 
should read
\begin{equation}
T\mathrm{d}S=\mathrm{d}E_{w} - 
\left( \Delta P \right)^{\rm sphere}_{\rm eff}\,\mathrm{d}V,
\label{therm_rel_sph}
\end{equation}
where 
\begin{eqnarray}
& & \left( \Delta P \right)_{\rm eff}^{\rm sphere}\,\mathrm{d}V = 
\left( \Delta P \right)_{\rm eff}\,\mathrm{d}V + \sigma_{\rm eff}\,\mathrm{d}A 
\nonumber \\
& & = \Delta P\,\mathrm{d}V + \sigma\,\mathrm{d}A - 
\left( \frac{\partial E_{m}}{\partial A} \right)_{V}\,\mathrm{d}A -
\left( \frac{\partial E_{m}}{\partial V} \right)_{A}\,\mathrm{d}V \nonumber \\
& & = \Delta P\,\mathrm{d}V + \sigma\,\mathrm{d}A - \mathrm{d}E_{m} 
\nonumber \\
& & = \left[ \Delta P - \left( \frac{\mathrm{d}E_{m}}{\mathrm{d}V} \right) +
\sigma \left( \frac{\mathrm{d}A}{\mathrm{d}V} \right) \right]\,\mathrm{d}V,
\label{p_sphere}
\end{eqnarray}
where in the last line the derivatives are not partial derivatives since for
a sphere $E_{m}=E_{m}(V)$ and $A=A(V)$. We may calculate these in a 
straightforward way. The mean curvature of a sphere of radius $r$ is $1/r$, 
and so from Eq.~(\ref{mem_ener}) 
\begin{equation}
E_{m} = 2 \pi \kappa \left( C_{0}r - 2 \right)^{2}\ \ \Rightarrow \ \
\frac{\mathrm{d}E_{m}}{\mathrm{d}V} = \frac{C_{0} \kappa}{r^2} 
\left(C_{0}r - 2 \right).
\label{E_sphere}
\end{equation}
Using $\mathrm{d}A/\mathrm{d}V = 2/r$, we find from Eqs.~(\ref{p_sphere}) 
and (\ref{E_sphere}) that
\begin{equation}
(\Delta P)_{\rm eff}^{\rm sphere} = \Delta P - 
\frac{C_{0}\kappa}{r^2}\left(C_{0}r - 2 \right) + \frac{2\sigma}{r}.
\label{Delta_p_eff_sphere}
\end{equation}
However for a sphere $J_{w} = L_{p} (\Delta P)_{\rm eff}$, and so in equilibrium 
when there is no flow of water, $J_{w}=0$, Eq.~(\ref{Delta_p_eff_sphere})
gives
\begin{equation}
\Delta P - \frac{C_{0}\kappa}{r^2}\left(C_{0}r - 2 \right) + \frac{2\sigma}{r} 
= 0.
\label{equilib_sphere}
\end{equation}
This is a standard result from the studies of spherical vesicles in 
equilibrium~\cite{ou87-89}, if we make allowance for the different sign 
conventions for the pressure difference and the lateral tension. In previous 
work~\cite{fan08,fan09}, we did not include the lateral tension in our 
description, and so our equilibrium result did not include the final term in 
Eq.~(\ref{equilib_sphere}). 

It should be emphasized that Eq.~(\ref{equilib_sphere}) is a consequence of
asking that the vesicle is in static equilibrium, so that in particular no
lipids are being added to the exterior surface, leading to no increase in the
surface area $A$. A condition for dynamic equilibrium can also be obtained.
This is a stationary state in which the vesicle remains turgid and grows like
a sphere. In this case, recalling that the area grows according to 
Eqs.~(\ref{growth_law}), one gets $\mathrm{d}V/\mathrm{d}t=2\pi\lambda r^{3}$. 
The spherical version of Eqs.~(\ref{J_w}) and (\ref{dV_by_dt}),
$\mathrm{d}V/\mathrm{d}t=AL_{p}\,(\Delta P)_{\rm eff}^{\rm sphere}$, therefore
gives
\begin{equation}
\frac{\lambda r}{2L_p} = \Delta P - 
\frac{C_{0}\kappa}{r^2}\left(C_{0}r - 2 \right) + \frac{2\sigma}{r},
\label{stationary_sphere}
\end{equation}
a result that could not be derived from a purely static description. 
Equation (\ref{stationary_sphere}) is Eq.~(21) of~\cite{fan08}, but with the
lateral tension now taken into account.

The inclusion of the lateral tension is even more important for axisymmetric
shapes of the kind we are considering here, because now $A$ is an independent 
variable. Since we are concerned with questions of stability, we will only
consider ellipsoids which differ in shape from the sphere very slightly. In 
this case the parameters $a$ and $c$ which describe the ellipsoid (see 
Eq.~(\ref{coords})) may be expressed as
\begin{equation}
a = R \left( 1 + a_{1} \epsilon \right)\,, \ \ c = R \left( 1 + c_{1} 
\epsilon \right)\,,
\label{a_and_c}
\end{equation}
where $\epsilon$ is a small quantity and $a_1$ and $c_1$ are numbers which 
characterize the shape of the ellipsoid: if $a_{1} > c_{1}$ it is oblate
and if $a_{1} < c_{1}$ it is prolate. The quantity $R$ reduces to the radius 
of the sphere as $\epsilon \to 0$, but care is required in its definition,
as we will discuss in more detail below. 

Using standard results\,\cite{bey87} and Eq.~(\ref{a_and_c}), it is 
straightforward to calculate the surface area and volume of the ellipsoid for 
small $\epsilon$. From the explicit forms given in the appendix of \cite{fan08}
it is found that
\begin{eqnarray}
A &=& 4\pi R^{2} \left[ 1 + \frac{2}{3} \left( 2a_{1} + c_{1} \right)\epsilon
+ {\cal O} \left( \epsilon^{2} \right) \right]\,, \nonumber \\
V &=& \frac{4}{3} \pi R^{3} \left[ 1 + \left( 2a_{1} + c_{1} \right)\epsilon
+ {\cal O} \left( \epsilon^{2} \right) \right]\,.
\label{A_and_V}
\end{eqnarray}
However, as we will now show, it is not consistent to assume that $R$ is
independent of $\epsilon$ if we assume a growth law of the 
form (\ref{growth_law}). To see this, we write the expression for $A$ given
in Eqs.~(\ref{A_and_V}) as $A=4\pi R^{2}\phi (\epsilon)$, where 
$\phi (\epsilon)$ is the expression in the square brackets. Then,
\begin{eqnarray}
\lambda A = \frac{\mathrm{d}A}{\mathrm{d}t} &=& 8\pi R \phi 
\frac{\mathrm{d}R}{\mathrm{d}t} + 
4\pi R^{2} \frac{\mathrm{d}\phi}{\mathrm{d}t} \nonumber \\
\Rightarrow \ \lambda &=& \frac{\mathrm{d} }{\mathrm{d}t} 
\left[ \ln R^{2}\phi \right],
\label{form_of_R}
\end{eqnarray} 
which implies that $R(t)=e^{\lambda t/2} \left[ \phi(t) \right]^{-1/2}$, up to
an overall multiplicative constant. Since $\phi$ is a function of $\epsilon$,
so is $R$. Therefore, for consistency, we cannot use $R$ when carrying out 
a perturbative expansion in $\epsilon$, since it contains hidden $\epsilon$ 
dependence. Instead we should use the radial variable
$r(t) = R(t) \left[ \phi(t) \right]^{1/2}$ which is $\epsilon-$independent. 
This is equivalent to determining $r$ through the condition $A=4\pi r^2$, for 
any axisymmetric ellipsoid. Clearly $r$ is the radius of the sphere which has 
the same surface area as the ellipsoid. 

The correct procedure to investigate the dynamics of vesicle growth 
perturbatively in $\epsilon$ is, therefore, to use closed form expressions from
\cite{fan08} and Eq.~(\ref{a_and_c}), to determine the results~(\ref{A_and_V}) 
to the required order, but then to set $A=4\pi r^2$. This can be inverted to
find $R(t) =r(t) \left[ \phi(t) \right]^{-1/2}$, allowing $V$ and $E_{m}$ to be
found as functions of $r$ and $\epsilon$. To first order, this procedure 
gives
\begin{equation}
\phi(t) = 1 + \frac{2}{3} \left( 2a_{1}+c_{1} \right) + 
{\cal O} \left( \epsilon^{2} \right),
\label{phi_to_first}
\end{equation}
leading to $V = (4\pi r^{3}/3) \left[ 1 + {\cal O} ( \epsilon^{2} ) \right]$. 
In fact, since $A=4\pi r^2$ exactly, the volume is directly related to 
the so-called reduced volume, defined by $v=6\sqrt{\pi}V/A^{3/2}$, by
\begin{equation}
V = \frac{4\pi r^3}{3} v.
\label{reduced_vol}
\end{equation}
Although $V$ has no term of order $\epsilon$ when expressed in terms of the 
variable $r$, it does turn out to have a term of order $\epsilon^2$:
\begin{equation}
v = 1 - \frac{4}{15} \left( a_{1}-c_{1} \right)^{2} \epsilon^{2} 
+ {\cal O} \left( \epsilon^{3} \right).
\label{v_to_second}
\end{equation}
As a check we note that $v<1$ for all cases except the sphere ($a_{1}=c_{1}$), 
as required. In our previous calculation~\cite{fan08} we used the variable
$R$, rather than $r$, which gave incorrect results for the coefficients in
the expansion in $\epsilon$. Correcting these by eliminating $R$ in favor of
$r$, makes previously cumbersome results look far more elegant. For example,
the energy of the membrane becomes
\begin{equation}
E_{m} (r,\epsilon) = E^{(0)}(r) + \alpha_{2} E^{(2)}(r) \epsilon^{2}   
+ {\cal O} \left( \epsilon^{3} \right),
\label{E_to_second}
\end{equation}
where 
\begin{equation}
E^{(0)}(r) = 2\pi \kappa \left( C_{0}r - 2 \right)^{2}, \ \
E^{(2)}(r) = \frac{8\pi \kappa}{3} \left( 6 - C_{0}r \right),
\label{E0_and_E2}
\end{equation}
and where
\begin{equation}\alpha_{2} = \frac{4}{15} \left( a_{1}-c_{1} \right)^{2}.
\label{alpha_2}
\end{equation}

A purely static analysis of the stability of an ellipsoidal vesicle would
compare the energy of the membrane (\ref{E_to_second}) to the energy of a 
spherical vesicle --- the same equation, but with 
$\epsilon = 0$ \cite{ou87-89}. The conclusion would be that the ellipsoid is 
the more stable if $E^{(2)}(r)<0$, i.e. if $C_{0}r > 6$. However there 
is no dynamics in this picture at all. To achieve a more physical description 
of the time evolution of the vesicle we utilize Eq.~(\ref{dV_by_dt}). To do 
this we first need to evaluate $(\partial E_{m}/\partial V)_{A}$ and 
$(\partial E_{m}/\partial A)_{V}$, but we cannot proceed directly, since we 
know $E_{m}=E_{m}(r,\epsilon)$ rather than $E_{m}=E_{m}(V,A)$. To circumvent 
this problem, suppose that we have inverted (in principle, not in practice) 
$V=V(r,\epsilon)$ to obtain $\epsilon=\epsilon(r,V)$. Then since 
$E_{m}=E_{m}(r,\epsilon(r,V))$, 
\begin{eqnarray}
& & \left( \frac{\partial E_{m}}{\partial V} \right)_{A} =
\left( \frac{\partial E_{m}}{\partial V} \right)_{r} =
\left( \frac{\partial E_{m}}{\partial \epsilon} \right)_{r}  
\left( \frac{\partial \epsilon}{\partial V} \right)_{r}, \nonumber \\
& & \left( \frac{\partial E_{m}}{\partial A} \right)_{V} =
\frac{1}{8\pi r}\,\left( \frac{\partial E_{m}}{\partial r} \right)_{V} =
\nonumber \\
& & \frac{1}{8\pi r}\,\left\{ \left( \frac{\partial E_{m}}
{\partial r} \right)_{\epsilon} + 
\left( \frac{\partial E_{m}}{\partial \epsilon} \right)_{r}  
\left( \frac{\partial \epsilon}{\partial r} \right)_{V} \right\}.
\label{chain_rule}
\end{eqnarray}
So we need only to calculate $(\partial \epsilon/\partial V)_{r}$ and   
$(\partial \epsilon/\partial r)_{V}$. These may be found from 
Eq.~(\ref{reduced_vol}) by noting that the reduced volume is a function only
of $\epsilon$. Then
\begin{equation}
v'(\epsilon) \left( \frac{\partial \epsilon}{\partial V} \right)_{r} =
\frac{1}{4\pi r^3/3}, \ \ 
v'(\epsilon) \left( \frac{\partial \epsilon}{\partial r} \right)_{V} =
- \frac{3}{r}\,\frac{V}{4\pi r^3/3},
\label{partial_derivs}
\end{equation}
where $v'(\epsilon)=\mathrm{d}v/\mathrm{d}\epsilon$. Substituting expressions 
(\ref{partial_derivs}) into Eqs.~(\ref{chain_rule}) gives
\begin{eqnarray}
& & \left( \frac{\partial E_{m}}{\partial V} \right)_{A} =
\frac{\left[ v'(\epsilon) \right]^{-1}}{4\pi r^{3}/3}\,
\left( \frac{\partial E_{m}}{\partial \epsilon} \right)_{r},
\nonumber \\
& & \left( \frac{\partial E_{m}}{\partial A} \right)_{V} =
\frac{1}{8\pi r}\,\left( \frac{\partial E_{m}}{\partial r} \right)_{\epsilon}
- \frac{3V}{2A}\,\left( \frac{\partial E_{m}}{\partial V} \right)_{A}.
\label{E_derivatives}
\end{eqnarray}
These two equations allow us to find $(\partial E_{m}/\partial V)_{A}$ and 
$(\partial E_{m}/\partial A)_{V}$ if we know $E_{m}$ and $V$ as functions of 
$r$ and $\epsilon$.

Using the results~(\ref{reduced_vol})-(\ref{alpha_2}) to second order in
$\epsilon$, the partial derivatives (\ref{E_derivatives}) are given by
\begin{eqnarray}
& & \left( \frac{\partial E_{m}}{\partial V} \right)_{A} = -
\left\{ \frac{1}{4\pi r^{3}/3} \right\} E^{(2)}(r) + 
{\cal O} \left( \epsilon \right),
\nonumber \\
& & \left( \frac{\partial E_{m}}{\partial A} \right)_{V} = \frac{1}{8\pi r} 
\left[ \frac{\mathrm{d}E^{(0)}}{\mathrm{d}r} + \frac{3}{r} E^{(2)}(r) \right] +
{\cal O} \left( \epsilon \right).
\label{deriv_2}
\end{eqnarray} 
It is important to note that while we have used the results for $E_{m}$ 
and $V$ correct to second order in $\epsilon$, we have only been able to 
calculate the required derivatives to zeroth order in $\epsilon$. The 
reason for this can be traced back to $v'(\epsilon)$ being of 
order $\epsilon$ and the differentiation of $E_{m}$ with respect to 
$\epsilon$ also giving an expression of order $\epsilon$. Together these 
reduce the powers of $\epsilon$ by $2$ in the calculation of 
$(\partial E_{m}/\partial V)_{A}$. The result can however be used to
check that we recover the previously derived form for the spherical vesicle 
in the limit $\epsilon \to 0$. Substituting Eqs.~(\ref{deriv_2}) into 
Eq.~(\ref{dV_by_dt}), and using $\mathrm{d}V/\mathrm{d}t=2\pi\lambda r^{3}$, 
valid for a sphere, one finds
\begin{eqnarray}
& & \frac{\lambda r}{2} = L_{p} \left\{ \Delta P + 
\frac{2\kappa}{r^3}\,\left( 6 - C_{0}r \right) \right\} \nonumber \\
& & + L_{\sigma} \left\{ \sigma - \frac{\kappa}{2 r^2} 
\left[ \left( C_{0}r \right)^{2} - 4C_{0}r + 12 \right] \right\}.
\label{stat_sphere}
\end{eqnarray}
This is identical to Eq.~(\ref{stationary_sphere}), already derived for the 
sphere provided that we make the identification $L_{\sigma} = 2L_{p}/r$. It 
should be emphasized that this identification is only being made to obtain 
the correct result in the spherical limit, bearing in mind that $L_{\sigma}$ 
is not defined in this case. 

\section{Stability}
\label{stable}

To determine if and when the shape of the vesicle starts to deviate from that
of a sphere, we calculate the time derivative of the reduced volume. This will
be zero ($v=1$) when the vesicle remains spherical, but will start to decrease 
as soon as it adopts another shape. From Eq.~(\ref{reduced_vol}), 
\begin{equation}
\frac{\mathrm{d}V}{\mathrm{d}t} = 4\pi r^{2} v\,\frac{\mathrm{d}r}{\mathrm{d}t}
+ \frac{4\pi r^3}{3}\,\frac{\mathrm{d}v}{\mathrm{d}t}.
\label{def_red_vol_change}
\end{equation}
The zeroth-order part of the first term on the right-hand side of 
Eq.~(\ref{def_red_vol_change}) has already been included in the stationary 
condition (\ref{stat_sphere}). At first order in $\epsilon$, the second term 
in this equation is the relevant one. To find the right-hand side of 
Eq.~(\ref{dV_by_dt}) to order $\epsilon$ it is necessary to find 
$(\partial E_{\rm m}/\partial V)_{A}$ to order $\epsilon^3$. This can be 
carried out in the same way as described above for the calculation to order
$\epsilon^2$, although the intermediate steps are sufficiently complicated 
that we used Mathematica~\cite{wolf}. Nevertheless, the final result is quite 
simple:
\begin{eqnarray}
- 2\alpha_{2}\epsilon\,\frac{\mathrm{d}\epsilon}{\mathrm{d}t} &=& 
\frac{4\kappa}{7r^4} \left(a_{1}-c_{1}\right) \nonumber \\
&\times& \left(5C_{0}r+6\right)\left[2L_{p}-rL_{\sigma}\right]\,
\epsilon + {\cal O} \left( \epsilon^{2} \right),
\label{result_1}
\end{eqnarray}
or defining a critical radius, $r_{c}$, by $r_{c}=2L_{p}/L_{\sigma}$, 
\begin{equation}
\alpha_{2} \frac{\mathrm{d}\epsilon}{\mathrm{d}t} = 
\frac{2\kappa L_{\sigma}}{7r^4} \left(a_{1}-c_{1}\right)\left(5C_{0}r+6\right)
\left(r-r_{c}\right) + {\cal O} \left( \epsilon \right).
\label{result_2}
\end{equation}

This stability condition differs from the one we had derived in our earlier
study of this question~\cite{fan08}, in that it predicts a linear growth 
away from the sphere, and not an exponential one. Since $\alpha_{2}>0$, if
the right-hand side of Eq.~(\ref{result_2}) is positive the sphere is 
linearly unstable, if it is negative it is linearly stable. We see that oblate
($a_{1}>c_{1}$) perturbations will destabilize a sphere if $r>r_c$, whereas
prolate ($a_{1}<c_{1}$) perturbations will destabilize a sphere if $r<r_c$.
The picture we have is that as the sphere grows, it is susceptible to 
fluctuations which give it a prolate ellipsoidal shape if it has a radius 
less than $r_c$, and susceptible to fluctuations which give it an oblate 
ellipsoidal shape if it has a radius greater than $r_c$.

How do these predictions compare with the currently available experimental
results? According to our predictions, spherical shapes should be unstable 
in a purely dynamical setting where the membrane surface grows due to the 
successive inclusion of lipid molecules and where the volume then increases 
due to the inflow of water. Under static conditions, however, the Helfrich 
energy functional implies that spherical vesicles are stable below a critical 
radius. It can be argued that the reason why spherical shapes are often seen 
in experiments~\cite{lui06} is a reflection of the specific experimental 
setting adopted. If the dynamical mechanisms of the type considered here are 
essentially negligible, then static effects are expected to prevail and, 
consequently, spheres are the energetically favored configurations. 

In~\cite{Nieh06}, the process of vesicle formation was investigated in a rich 
and dynamic phospholipid mixture. Dynamical light scattering and transmission 
electron microscopy experiments were performed to resolve the vesicles' shape 
and quantify their associated sizes. The observed vesicles were relatively 
large (hydrodynamic radii $\simeq 200$~nm or $> 500$~nm) and corresponded to 
either oblate ellipsoids or triaxial ellipsoids. The fact that a significant 
fraction of the (giant) vesicles belonged to the oblate ellipsoids family is 
consistent with the prediction from our analysis. 

Clearly, there are number of other effects (e.g. temperature fluctuations) 
which are implicated in the above mentioned experiments and not included 
in our modeling efforts. We also need to consider more generic shapes, 
including triaxial ellipsoids. The probability that perturbations 
are axisymmetric are presumably small, nevertheless, oblate ellipsoids are 
observed in experiments where the dynamics plays a role and the vesicles are 
sufficiently large. This is for instance the case in~\cite{Ben98,Pen01}, 
where initial spherical vesicles are experimentally shown to deform into 
oblate ellipsoids, rather then prolate ones, in the presence of osmotic driving 
pressures. This is a distinctive feature which is not captured by any static 
approaches to the problem of the morphology of vesicles and one that could 
perhaps be explained by a crucial interplay between shape and dynamics.

\section{Conclusions}
\label{conclude}

In this paper we have extended and corrected earlier work on the growth 
and stability of vesicles. The errors in our previous paper~\cite{fan08} 
were firstly an inappropriate choice of characteristic size for the
ellipsoidal vesicle ($R$ rather than $r$). This was a subtle point relating 
to the fact that assuming the radius to be independent of the degree of 
deformation $\epsilon$ was not compatible with the growth law we chose. The 
second error was one of omission: we did not include the lateral tension in
our original analysis. Finally, there was a simple typographical error in
Eqs.~(A9) of~\cite{fan08}. While correcting the first point would not have 
changed the conclusions to any great extent, adding the surface tension does:
the prediction is now that smaller vesicles should tend to be prolate and 
larger vesicles should tend to be oblate. 

There are several caveats that we should make in regard to this prediction. 
Foremost among these is that the model we have adopted is very simple, and 
even though it is the one frequently used in the literature, we should keep
this constantly in mind. It should be possible to extend these results 
to the ADE~\cite{miao94} model, which is slightly more realistic. It would be 
interesting to see if the predictions are changed in any way. Another caveat 
is that we have assumed that the sphere becomes unstable to an ellipse 
defined by Eqs.~(\ref{coords}). Although this is what is found from 
investigations using variational techniques~\cite{ou87-89}, it would be more 
consistent not to assign a specific functional form to the shape of the
membrane. This is currently being investigated; the analysis is considerably 
more complicated and the results will be presented elsewhere~\cite{mor10}. 
Finally, LNET for this problem is very underdeveloped, and rather complex, 
with plenty of scope for pitfalls. This will also be more extensively discussed 
elsewhere~\cite{mor10}.

Even given all these caveats, we still believe that the current work is a 
significant step forward in understanding vesicle growth and shape changes. 
Virtually all previous work was static and made much less contact with the 
physics of the problem than the dynamic approach adopted here. What work
did exist on the dynamics, including our own, has errors or inconsistencies 
which we hope that we have rectified. What is required most of all are
more experiments in order to guide the theoretical development, ruling out
and supporting the various theoretical approaches. We hope that the current 
work stimulates the carrying out of such experiments in the future. 
 
\acknowledgments

RGM wishes to thank the EPSRC (UK) for the award of a postgraduate 
studentship.


\end{document}